\documentclass[aps,prb,twocolumn,superscriptaddress]{revtex4-1}
\usepackage{graphicx}
\usepackage{latexsym}
\usepackage{amssymb}
\usepackage{amsmath}
\usepackage{amsfonts}
\usepackage{upgreek}
\usepackage{subfigure}
\usepackage{bm}
\usepackage{bbold}
\usepackage[unicode=true,
 bookmarks=false,
 breaklinks=false,pdfborder={0 0 1},backref=false,colorlinks=true]
 {hyperref}
\hypersetup{
 linkcolor=[rgb]{0,0,1},citecolor=[rgb]{0,0,1},urlcolor=[rgb]{0,0,1}}
\usepackage{multirow}
\usepackage{color}
\usepackage{comment}
\usepackage{xcolor}
\usepackage{soul}
\usepackage{tikz}
\newcommand{\overbar}[1]{\mkern 1.5mu\overline{\mkern-1.5mu#1\mkern-1.5mu}\mkern 1.5mu}
\newcommand{\angstrom}{\textup{\AA}}
\usepackage{braket}
\newcommand{\Apix}{%
        \setcounter{table}{0}
        \renewcommand{\thetable}{A\arabic{table}}%
        \setcounter{figure}{0}
        \renewcommand{\thefigure}{A\arabic{figure}}%
        \setcounter{equation}{0}
        \renewcommand{\theequation}{A\arabic{equation}}
        \setcounter{section}{0}
        \renewcommand{\thesection}{Appendix \Alph{section}}
     }

\usepackage{standalone}

\begin{document}

\title{Plasmons in Two-Dimensional Topological Insulators}

\author{Henning Schl\"omer}
\affiliation{Institute for Theoretical Solid State Physics, RWTH Aachen University, 52056 Aachen, Germany}
\author{Zhihao Jiang}
\affiliation{Department of Physics and Astronomy, University of Southern California, Los Angeles, CA 90089-0484}
\author{Stephan Haas}
\affiliation{Department of Physics and Astronomy, University of Southern California, Los Angeles, CA 90089-0484}

\date{\today}
\begin{abstract}
We analyze collective excitations in models of two-dimensional topological insulators using the random phase approximation. In a two-dimensional extension of the Su-Schrieffer-Heeger model, edge plasmonic excitations with induced charge-density distributions localized at the boundaries of the system are found in the topologically non-trivial phase, dispersing similarly as one-dimensional bulk plasmons in the conventional Su-Schrieffer-Heeger chain. For two-dimensional bulk collective modes, we reveal regimes of enhanced inter-band wave function correlations, leading to characteristic hardening and softening of inter- and intra-band bulk plasmonic branches, respectively. In the two-dimensional Haldane Chern insulator model, chiral, uni-directional edge plasmons in nano-ribbon architectures are observed, which can be characterized by an effective Coulomb interaction cross section. Bulk collective excitations in the two-dimensional Haldane model are shown to be originated by single-particle band structure details in different topological phases.
\end{abstract}
\maketitle

\section{Introduction}
\label{sec:intro}
Non-interacting models of topological insulators have been studied to a great extend in recent years, and have become one of the most active and rapidly growing research areas in condensed matter physics~\cite{Ando2013,Hsieh2008,Chang2013,Sato2016,Hasan2010,Qi2011,Fu2008,Laughlin1981,Ando1974,Thouless1982,Laughlin1983,Hirsch1999,Chen2009,Murakami2004,Kane2005,Kane2005_2,Bernevig2006, Bernevig2006_2,Konig2007,Moore2007,Fu2007,Fu2007_2,Zhang2001,Qi2008,Nomura2011,Kim2012,Tanaka2012,Haldane2004}. However, effects of electron-electron correlations in topological insulators have not yet received equal attention. Nonetheless, measurements of collective electronic excitations arising from long-range Coulomb interaction are experimentally more accessible than probing directly single particle states, as it has been demonstrated e.g. for thin micro-ribbon arrays of single layer graphene~\cite{Ju2011} and three-dimensional topological insulator thin films ($\text{Bi}_{2}\text{Se}_{3}$)~\cite{Pietro2013,Pietro2020} using infrared spectroscopy. Recently, plasmons in the one-dimensional (1D) Su-Schrieffer-Heeger (SSH) model were analyzed in real space, where it was found that single particle edge states appearing in the non-trivial phase lead to strongly localized plasmon charge distributions~\cite{Zhihao1}. 
Furthermore, bilayer architectures composed of two unhybridized but Coulomb coupled massless Dirac electron systems,  realized on surfaces of three-dimensional (3D) topological insulators, have  been studied theoretically~\cite{Rosario2012}. Here, the typically large bulk dielectric screening was observed to lock the low-frequency plasmon modes at energies above the particle-hole (p-h) continuum. In magnetically doped thin films of 3D topological insulators, band-inversions were shown to enhance inter-band correlations, which in turn lead to the appearance of inter-band plasmonic responses in certain topological phases \cite{Zhang2017}.

In this paper, we report results for collective plasmonic excitations in two-dimensional (2D) topological insulators. Specifically, we investigate plasmonic responses in real and reciprocal space, capturing their bulk and surface properties, respectively. For the 2D extension of the SSH model, we find that topological mid-gap edge states open a quasi-1D plasmonic channel, which disperse like their 1D SSH bulk plasmon analogues. In the 2D Haldane model, chiral, uni-directional, quasi-1D plasmons emerge in the topologically non-trivial phase, whose dispersion is characterized by the the quasi-1D Coulomb cross section that effectively describes inter-band screening effects. In the 2D SSH model, we further identify two distinct regimes, namely dimerized (D) and anti-dimerized (AD), which depend on the choice of hopping parameters and on the real space lattice modulation and significantly influence the bulk plasmonic response. We observe high-energy plasmons well above the p-h continua in the AD phase, which are hence intrinsically undamped. We envision an experimental setup on a momentum-space lattice \cite{Meier2016, Meier2016_2}, allowing for manipulation of the hopping parameters and real space lattice structure, thus enabling access to the different plasmonic excitation spectra, which can be observed via electronic energy loss spectroscopy \cite{Eberlein2008} or electromagnetic radiation combined with sub-wavelength grated surface probes~\cite{Ju2011,Pietro2013}. For the detection of high-energy, intrinsically undamped plasmons in the case of the 2D SSH model, we propose a method analogous to the concepts introduced in~\cite{Lewandowski2019}, where it was argued that typical speckle patterns produced by elastic scattering processes can be observed via spatial near-field imaging in regimes where Landau damping is quenched.

The paper is organized as follows. In Sec~\ref{sec:methods}, we introduce the methods used throughout this work. In Sec.~\ref{sec:2Dssh}, bulk and surface plasmons in the 2D SSH model are analyzed. In the Haldane model, we calculate collective excitations in nanoribbon- and bulk-materials in Sec.~\ref{sec:Haldane}, after which we conclude our findings in Sec.~\ref{sec:Conc}.
%
%
%
%
\section{Methods}
\label{sec:methods}
We account for long-range Coulomb interactions via the random-phase approximation (RPA). The complex dielectric function, whose nodes yield a diverging dynamical response to an external electric perturbation, is given by
\begin{equation}
\epsilon(\omega,\mathbf{q}) = 1- V_{\mathbf{q}} \Pi^0(\omega, \mathbf{q}),
\label{eq:RPA}
\end{equation}
where $V_{q}= 2\pi e^2/\kappa q$ is the Fourier transformed Coulomb potential in two spatial dimensions with background dielectric screening $\kappa$, and  $\Pi^0(\omega, \mathbf{q})$ denotes the bare polarization function. For a system with $n$ sites per unit cell and band index $l=1...n$, the bare polarization bubble can be evaluated~\cite{Tsuneya2006},
\begin{flalign}
\Pi^0(\omega, \mathbf{q}) = \frac{g_s}{V} \sum_{\mathbf{k},l, l'} \frac{n_F(E_{\mathbf{k},l}) - n_F(E_{\mathbf{k+q},l'})}{\omega + i\eta + E_{\mathbf{k},l} - E_{\mathbf{k+q},l'}} F_{\mathbf{k}, \mathbf{k+q}}^{l l'}. &&
\raisetag{2.5em}
\label{eq:Pi0}
\end{flalign}
Here, $g_s=2$ is the spin degeneracy factor, $V$ is the volume of the unit cell, $\sum_{\mathbf{k}}$ runs over the first Brillouin zone (BZ), $E_{\mathbf{k},l}$ are the single particle states, $n_F(x) = (1+\exp[-(x-\mu)/k_B T])^{-1}$ is the Fermi-Dirac distribution function at chemical potential $\mu$, and $F_{\mathbf{k}, \mathbf{k+q}}^{l l'}$ is the overlap function of the corresponding pseudo-spinors \footnote{We here choose a form of the momentum space Hamiltonian that fulfills the discretization of the Bloch theorem, the discrete variable being the internal degree of freedom (i.e., the sites in the unit cell). For the other conventional form of momentum Hamiltonians, which differs from the discrete Bloch theorem form by a site-dependent gauge transformation, the overlap functions would read $F_{\mathbf{k}, \mathbf{k+q}}^{l  l'} = |\braket{\Psi_{\mathbf{k}, l}|e^{i\mathbf{q}\cdot \mathbf{r}|\Psi_{\mathbf{k+q},l'}}}|^2$, where the real space operator takes the value of the real space coordinate of each entry.},
\begin{equation}
    F_{\mathbf{k}, \mathbf{k+q}}^{l  l'} = |\braket{\Psi_{\mathbf{k}, l}|\Psi_{\mathbf{k+q},l'}}|^2.
\end{equation}
The plasmon dispersion can then be extracted from the electronic energy loss spectrum (EELS) given by $\text{EELS}(\omega,\mathbf{q}) = -\text{Im} 1/\epsilon(\omega, \mathbf{q})$.
In real space, assuming the system consists of $M$ unit cells and thus $N=n M$ total sites, we calculate the $N \times N$ response matrix $\boldsymbol{\epsilon}(\omega) = \mathbb{1} - \mathbf{V} \boldsymbol{\Pi}^0$, where $\mathbf{V}$ is the real space Coulomb interaction matrix,
\begin{equation}
    \mathbf{V}_{ab} = \begin{cases} 
    e^2/\kappa|\mathbf{r}_a - \mathbf{r}_b| \qquad a\neq b \\
    U_0/\kappa \qquad \qquad \quad \,\,\, a=b
    \end{cases}
\end{equation}
with $U_0 = 17.38$ eV~\footnote{$U_0 = \int d\mathbf{r} d\mathbf{r}' e^2 |\phi(\mathbf{r}, \sigma^2)|^2 |\phi(\mathbf{r}', \sigma^2)|^2/|\mathbf{r}-\mathbf{r}'|$, where $\phi(\mathbf{r}, \sigma^2)$ is the 2D Gaussian distribution with standard deviation $\sigma= \delta = 1 \angstrom$.} and
\begin{equation}
    [\boldsymbol{\Pi}^0]_{ab} = g_s \sum_{i,j} \frac{n_F(E_{i}) - n_F(E_{j})}{E_{i} - E_{j} - \omega - i\eta} \psi_{ia}^* \psi_{ib} \psi_{jb}^* \psi_{ja},
\end{equation}
the bare polarization function in real space. Here, $E_i$, $n_F(E_i)$ and $\psi_{ia}$ are the $i$th electronic eigenenergy, the corresponding Fermi function at chemical potential $\mu$ and the wave function coefficient of tight binding orbital $a$, respectively. The real space non-interacting density response is efficiently calculated using a Green's function approach based on fast Fourier transforms~\cite{Sukosin2012,honet2020semiempirical, Shishkin2006}. We then extract the electronic energy loss function by choosing the eigenvalue $\epsilon_n(\omega)$ and eigenvector $\mathbf{v}_{\text{max}}$ of $\boldsymbol{\epsilon}(\omega)$ such that $\text{EELS}(\omega) = -\text{Im}1/\epsilon_n(\omega)$ is maximized~\cite{Westerhout2018, Wang2015}. For a qualitative picture of the induced charge distribution $\boldsymbol{\rho}(\omega) \propto \boldsymbol{\Pi}^0 \mathbf{v}_{\text{max}}$ of the plasmon modes, we approximate the tight binding orbital around site $a$ as a 2D Gaussian distribution $\phi_a(\mathbf{r}, \sigma^2)$ with variance $\sigma^2$ and transform the charge density distribution into the $\mathbf{r}$-space representation via $\rho(\omega,\mathbf{r}) = \sum_a \boldsymbol{\rho}_{a}(\omega) \phi_a(\mathbf{r}, \sigma^2)$. Throughout our computations, we set $\hbar=1$, $T=0$ K, and use numerical broadenings $\eta=0.01$ eV and $\eta=0.08$ eV in momentum and real space, respectively.

%
%
%
%
\section{2D Su-Schrieffer-Heeger Model}
\label{sec:2Dssh}
\begin{figure}
\centering
\includegraphics[width=1.0\linewidth]{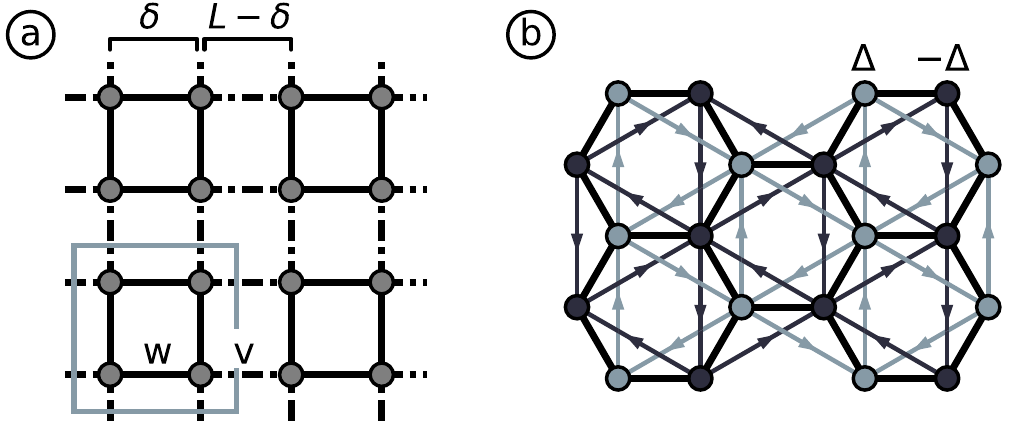}
\caption{Topological insulator models analyzed in this article. (a) Two-dimensional extension of the SSH model. Each unit cell (light blue solid line) includes four sites, with intra-cell nearest neighbor hopping $w$ and inter-cell tunneling $v$. Nearest neighbor distances within each unit cell plaquette are given by $\delta$, whereas inter-cell neighbors measure a distance of $L-\delta$. (b) Haldane model, a Chern insulator on the honeycomb lattice consisting of nearest neighbor (black solid line) and complex next-nearest neighbor hoppings (light and dark blue solid lines) sharing the same chirality (indicated by arrows). A staggered sublattice on-site potential $\Delta$ (-$\Delta$) on A (B) sites is further present.}
\label{fig:models}
\end{figure}

We start by analyzing a two-dimensional extension of the SSH model~\cite{Liu2017, Obana2019}, i.e., a square lattice with $2\times 2$ sites per unit cell and intra-cell (inter-cell) hopping $w$ ($v$), Fig.~\ref{fig:models} (a).
The $4\times 4$ Hamiltonian in reciprocal space has the entries 
\begin{equation}
\begin{split}
    &\mathcal{H}_{12} = \mathcal{H}_{34} = w \exp\{ik_x \delta\} + v \exp\{-ik_x (L-\delta)\} \\ &\mathcal{H}_{13} = \mathcal{H}_{24} = w \exp\{ik_y \delta\} + v \exp\{-ik_y (L-\delta)\},
\end{split}
\label{eq:2dssh_Ham}
\end{equation} 
with their corresponding complex conjugate partners at transposed matrix elements, $L^2$ the surface of the unit cell, and $\delta$ ($L-\delta$) the intra-cell (inter-cell) nearest neighbor distance. Having time reversal (TR) and inversion symmetry, the Berry curvature vanishes throughout the entire BZ, except at $C_{4v}$ invariant points $|k_x|=|k_y|$, where oscillating divergences appear due to the degeneracy of energy bands. These, however, integrate to zero and thus result in a vanishing Chern number. Nevertheless, a non-trivial topological classification arises through a finite 2D Zak-phase, resulting in a fractional wave polarization and topological edge states for $w<v$~\cite{Liu2017}, hence resembling its analogue in one spatial dimension~\cite{Su1980, Heeger1988}. 
\begin{figure*}
\centering
\hspace*{-0.4cm}
\begin{subfigure}
  \centering
  \includegraphics[width=9.396cm]{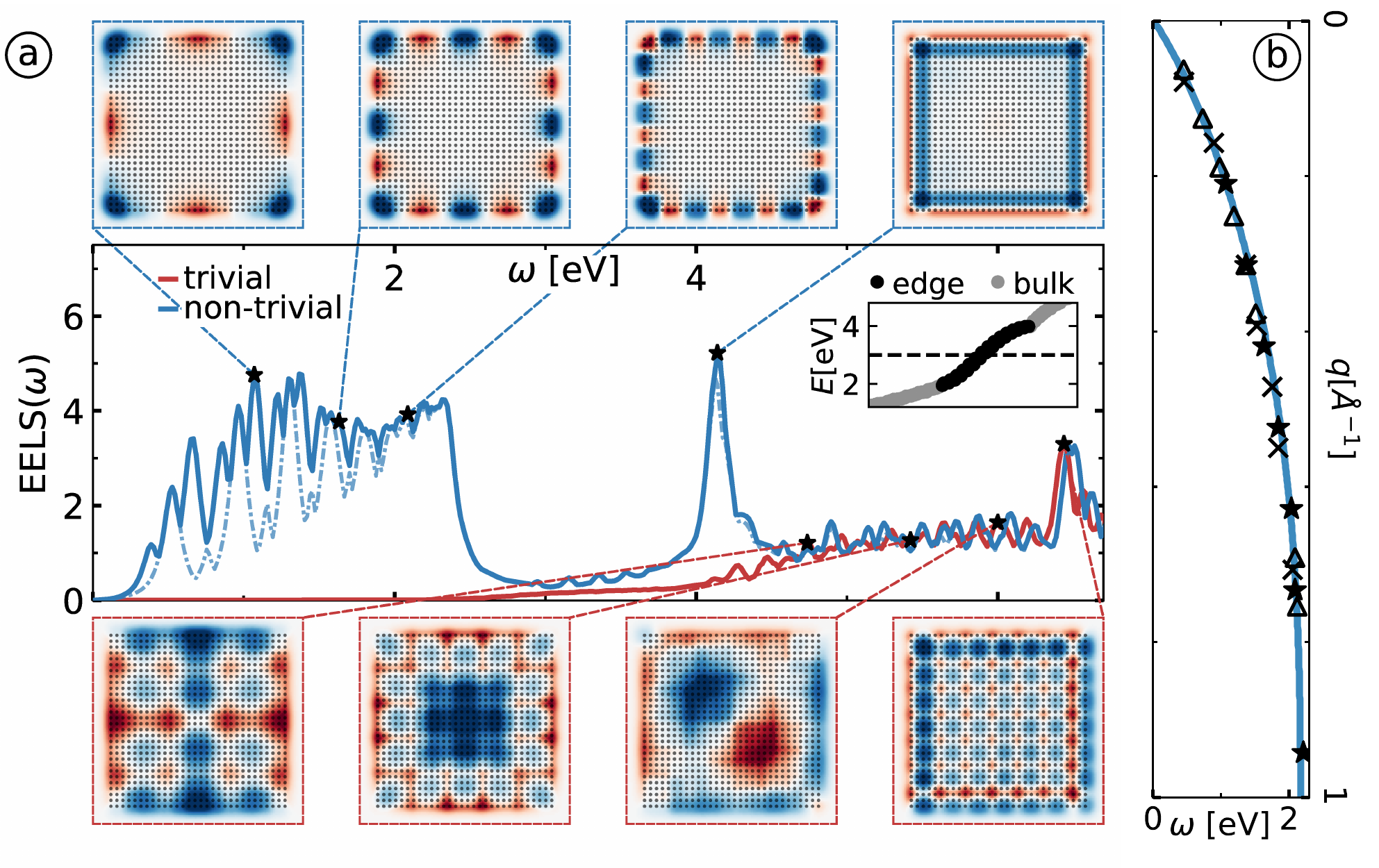}
\end{subfigure}%
\begin{subfigure}
  \centering
  \includegraphics[width=8.176cm]{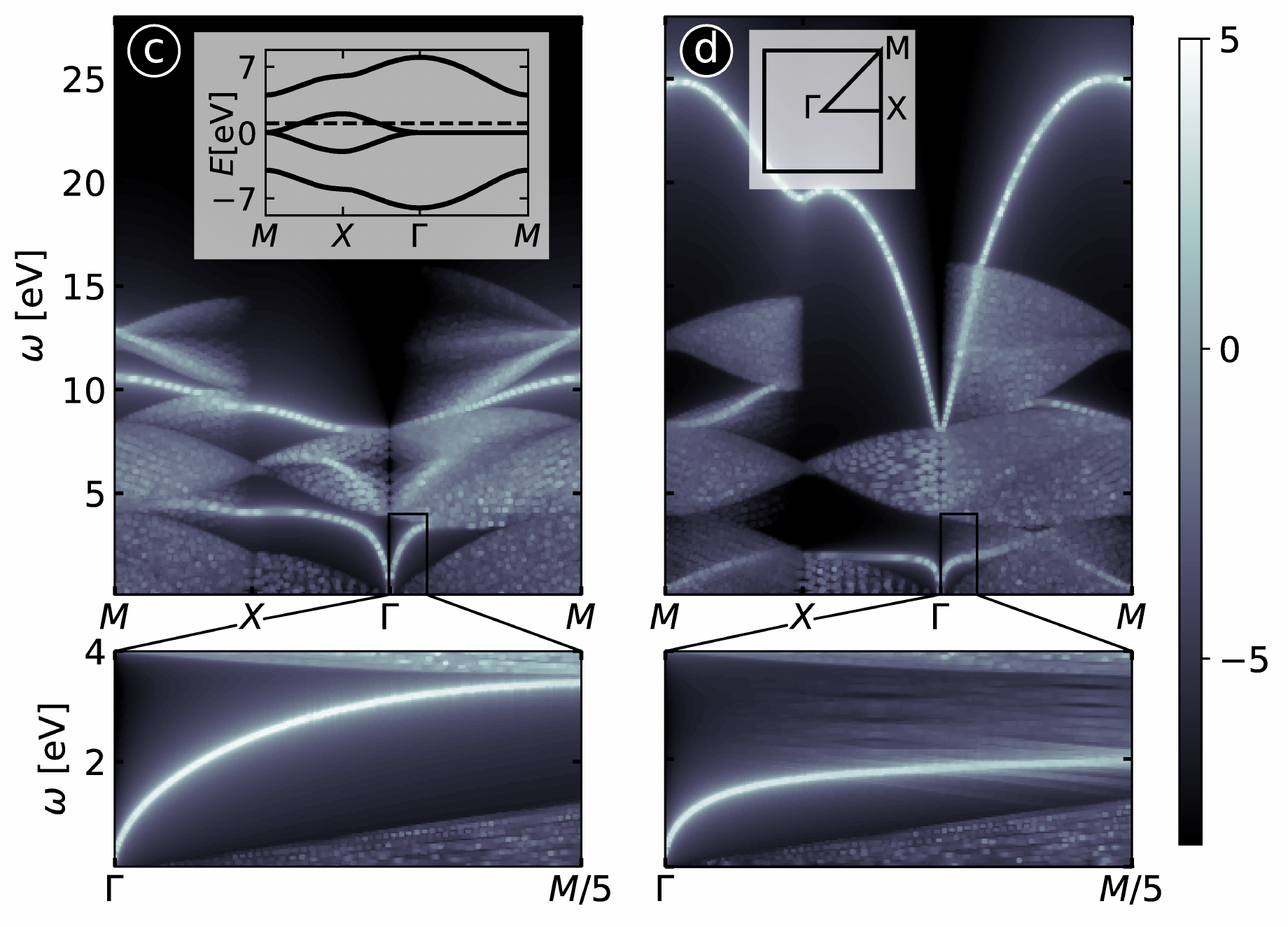}
\end{subfigure}%
\caption{(a) $\textit{Real space:}$ $\text{EELS}(\omega)$ and real space plasmonic charge distributions for a $30\times 30$ site system with uniform atomic distance $\delta= L-\delta =1$ $\angstrom$. Center plot: $\text{EELS}(\omega)$ for the trivial ($w=1$ eV, $v=3$ eV, red solid line) and non-trivial ($w=3$ eV, $v=1$ eV, blue solid line) phase. The second leading eigenvalue is also shown for the non-trivial phase (blue dashed line). The inset illustrates the mid-gap edge states (black dots) in the upper band gap appearing in addition to bulk states (grey dots) in the non-trivial phase, as well as the chemical potential $\mu=3$ eV (black dashed line). Top and bottom row: charge distributions $\rho(\omega,\mathbf{r})$ of chosen peaks for the non-trivial and trivial regime, respectively. A background dielectric screening $\kappa=2.5$ is used and $\sigma = 1$ $\angstrom$. (b) Edge plasmon dispersion read off from the real space calculation for system sizes $30 \times 30$ (stars, see (a)), $40 \times 20$ (crosses), $50 \times 50$ (triangles), and the 1D SSH bulk plasmon dispersion in momentum space using identical parameters as in (a) and a Coulomb cross section $\sigma_C = 1.1 \delta$. (a)\&(b) $\textit{Momentum space: }$ Electron energy loss spectra ($\text{log}[\text{EELS}(\omega,\mathbf{q})]$) of the 2D SSH model for (c) the dimerized ($w=t_s=3$ eV, $v=t_l=1$ eV), and (d) the anti-dimerized regime ($w=t_s=1$ eV, $v=t_l=3$ eV). Inset (a): single-particle energy dispersion (solid lines) and chemical potential $\mu=1$ eV (dashed line). Lower plots show zooms of the low energy plasmon modes. The unit cell size is set to $L=2$ $\angstrom$, and $\delta=0.1 L$.}%
    \label{fig:2dssh_plasmons}
\end{figure*}

Fig.~\ref{fig:2dssh_plasmons} (a) depicts the real space results in a finite system of $30\times 30$ sites on a uniformly spaced square lattice, i.e., $\delta=L/2$. In order to energetically decouple the contributions coming from the bulk and mid-gap topological edge states appearing in the non-trivial phase, we here tune the chemical potential to lie inside the upper bulk band gap, as shown in the inset of Fig.~\ref{fig:2dssh_plasmons}~(a). Furthermore, a background dielectric constant of $\kappa=2.5$ is chosen.
Examining the electronic energy loss spectrum, using the leading eigenvalue method, we observe that for plasmonic energies $\omega \lesssim 2.5$ eV, a collective excitation continuum arises in the topologically non-trivial phase, with quasi-1D induced charge modulations localized at the boundaries of the slab (upper row in Fig.~\ref{fig:2dssh_plasmons}~(a)). Bulk plasmons ($\omega \gtrsim 4.5$ eV), on the other hand, have similar loss spectra in both phases and are characterized by induced charge distributions delocalized throughout the bulk of the system, cf. lower row of Fig.~\ref{fig:2dssh_plasmons}~(a). These similarities in opposing phases are expected for $\delta=L/2$, since in that particular case the Hamiltonian, Eq.~\eqref{eq:2dssh_Ham}, entering the polarization function is invariant under the transformation $w \leftrightarrow v$.
The appearance of localized collective excitations in the non-trivial phase shows how single particle transitions involving the mid-gap topological edge states open a quasi-1D plasmonic channel, thus generalizing the results presented in Ref.~\cite{Zhihao1} to higher dimensions.

Analyzing the wavelength of the plasmon waves appearing in the localized plasmonic spectrum, we extract their dispersion $\omega(q)$ for various system shapes and sizes. We further numerically evaluate the RPA bulk plasmon dispersion of the 1D SSH model in reciprocal space with identical hopping and environmental parameters and using the quasi-1D Coulomb interaction in Fourier space, $V(q) = 2e^2 K_0(\sigma_C |q|)/\kappa$. Here, $K_0(\cdot)$ is the zeroth modified Bessel function of second kind, and $\sigma_C$ is the effective cross section of the 1D material embedded in 3D space \cite{del1992}. We find that the localized edge plasmon dispersion of the 2D system matches its 1D analogue, with a Coulomb cross section of $\sigma_C = 1.1 \delta$, {\it independent} of the system size and shape, as illustrated in Fig.~\ref{fig:2dssh_plasmons}~(d). This highlights the emergence of quasi-1D collective physics in the 2D SSH model in topologically non-trivial phases, with the plasmons being confined to a region of size $\approx \delta$. Furthermore, the independence of the edge plasmon dispersion on the system size and shape suggests a full decoupling of edge and bulk modes, whereby bulk screening effects do not influence the quasi-1D collective excitations. Note that certain similarities of the 1D SSH and quasi-1D edge physics in the 2D model are expected, as the non-trivial edge modes in 2D SSH nanoribbons disperse identically to bulk 1D SSH bands, and the emerging edge states are localized throughout the entire BZ, hence being decoupled from the bulk modes~\cite{Liu2017, Obana2019}.

The localized plasmonic peak at the lower border of the bulk continuum ($\omega \approx 4.2$ eV), additionally appearing in the non-trivial phase, shows a charge modulation perpendicular instead of parallel to the surface, which leads to a larger generated Coulomb energy and hence separates it from the edge continuum.
We further find that charge responses that break $C_{4v}$ symmetry are degenerate, the induced charge-density distributions differing from each other only by a $C_{4v}$ symmetry operation, as indicated by the second largest $\text{EELS}(\omega)$ in Fig.~\ref{fig:2dssh_plasmons}(a). 

Let us now turn to bulk collective physics in the 2D SSH model. As already mentioned in the discussion above, bulk plasmons do not differ in opposing topological phases for a uniformly spaced square lattice. However, when additionally tuning the intra-cell atomic distances, interesting inversion effects can be observed, which in turn strongly influence the bulk plasmonic response. In our calculations, we fix the intra-cell nearest neighbor distance to $\delta=0.1 L$. For clarity, we  refrain here from using the terms topologically trivial and non-trivial, and instead use the following terminology: if the tunneling amplitude $t_s$ associated with bond of length $\text{min}\{\delta, L-\delta\}$ is larger [resp. smaller] than the hopping $t_l$ corresponding to two atomic sites separated by $\text{max}\{\delta, L-\delta\}$, the phase is referred to as dimerized (D) [resp. anti-dimerized (AD)]. For our choice of $\delta$, phase D (AD) corresponds to the topologically trivial (non-trivial) phase of the 2D SSH model. A change between the D$\leftrightarrow$AD regimes can then either be induced by a topological phase transition $t_s \leftrightarrow t_l$ or by a change $\delta \leftrightarrow L-\delta$. Figs.~\ref{fig:2dssh_plasmons} (a) and (b) show the electronic loss functions of the model in the dimerized and anti-dimerized phase in momentum space, respectively. The inset of Fig.~\ref{fig:2dssh_plasmons} (a) illustrates the four single-particle energy bands through high-symmetry points of the BZ, referred to as $s,p_x,p_y$ and $d_{xy}$ from lowest to highest energy. The chemical potential is chosen to lie inside the $p_y$ band, which opens both intra- and inter-band polarization channels and thus enables us to compute both gapless and gapped plasmonic modes. Focusing first on the dimerized regime, Fig.~\ref{fig:2dssh_plasmons} (a), we see that the energetically lowest plasmon branch governed by $p_y$ intra-band transitions is characterized by a steep dispersion, entering the p-h continua for small momentum transfers close to the $\Gamma$ point when moving towards $M$. Inter-band plasmonic excitations, on the other hand, feature flat energy dispersions, hybridizing with the p-h continua and hence being Landau damped throughout the BZ. 

When tuning the bulk into the AD phase, the resulting collective excitation spectra are in stark contrast to the D regime. The gapless plasmonic intra-band mode is softened (i.e., red-shift), reaching up to only about half of the maximum energy compared to its analogue in the D phase, and then hybridizing with the p-h continuum. For high energy, gapped plasmons we observe the opposite, namely a strong hardening (i.e., blue-shift) of the inter-band modes. This sharp increase of collective excitation energies leads to intrinsically undamped plasmonic modes already for small momentum transfers $q$, whereby Landau damping via p-h excitations is almost entirely quenched. Note that the enhancement of plasmonic energies into intrinsically undamped regimes is reminiscent of what was found in flat-band Hamiltonians such as twisted-bilayer graphene due to large fine structure constants~\cite{Lewandowski2019}, which is, however, of a different physical nature.

Sharing identical single-particle energy dispersions in both phases due to the sublattice symmetry, the only factor leading to different plasmonic excitation spectra lies in the overlap of the wave functions $F_{\mathbf{k}, \mathbf{k+q}}^{l l'}$, which store the real space and topological properties of the system and hence influence the collective modes accordingly. Indeed, we find that inter-band (intra-band) correlations are greatly enhanced (suppressed) in the AD regime. This, in turn, leads to larger (smaller) screening of the effective electron interaction, which ultimately results in the observed softening (hardening) of the gapless (gapped) plasmon branches.~\footnote{The overlap functions, entering via the Coulomb interaction matrix elements, also affect the p-h continua given by peaks of $\text{Im}\Pi^0(\omega,\mathbf{q})$. This results in single-particle transitions to be suppressed along certain momentum transfer paths in the BZ. Prominent suppression occurs especially along the $\Gamma-X$ and $\Gamma-M$ directions, the latter being less noticeable in the AD regime due to larger inter-band correlations as discussed in the main text.} All 16 overlap functions for paths along high-symmetry points of the BZ in both phases are shown in~\ref{sec:ApxA}.
Note that, for the standard tight-binding insulator model on a 2D square lattice, only marginal differences are observed in the two phases, such that the imbalance of hopping amplitudes is necessary to achieve the observed effects. From an experimental point of view, we propose a 2D array of atoms trapped in a periodic potential well, where the tunability of the barrier width and height enables the realization of the more uncommon anti-dimerized phase. This, as well as a more fundamental analysis of the real space effects on plasmonic spectra in 1D topological insulator models, is addressed in an upcoming work~\cite{Zhihao2}. We would like to stress that the observed tunability of bulk plasmons should not be considered as a \textit{generic effect} originating from its non-trivial topology, but rather as an additional property of the topological tight-binding model at hand.


%
%
%
%
\section{Chern Insulator on the Honeycomb Lattice}
\label{sec:Haldane}
We now analyze plasmonic excitation spectra in the Haldane model~\cite{Haldane1988}, the most prominent example of a quantum anomalous Hall (QAH) insulator~\cite{Liu2016} featuring the quantum Hall effect with vanishing net-magnetic flux. It is a tight binding model on a honeycomb lattice, allowing for real nearest neighbor hoppings $t$, complex (TR symmetry breaking) next-nearest neighbor tunneling terms $it'$ of uniform chirality, as well as a staggered (sub-lattice symmetry breaking) on-site potential $\Delta$ opening a gap. Fig.~\ref{fig:models} (b) illustrates the Haldane model. The Hamiltonian in momentum space takes the form
\begin{equation}
    \mathcal{H}(\mathbf{k}) =\mathcal{H}_G + \Big( \Delta + 2 t^{\prime} \sum_i \sin(\mathbf{k}\cdot \mathbf{b}_i) \Big) \sigma_z,
\label{eq:H_haldane}
\end{equation}
where $\mathcal{H}_G$ is the nearest-neighbor tight binding Hamiltonian for graphene
\footnote{The nearest neighbor tight-binding Hamiltonian for graphene is given by 
\begin{equation*}
\begin{split}
    \mathcal{H}_G &= \left( \begin{array}{cc}
         0&h(\mathbf{k})  \\
         h^*(\mathbf{k})&0 
    \end{array} \right) \\
    &= t \sum_i \Big( \sigma_x \cos(\mathbf{k}\cdot \mathbf{a}_i) - \sigma_y \sin(\mathbf{k}\cdot \mathbf{a}_i) \Big),
\end{split}
\label{eq:H_graphene}
\end{equation*}
where $h(\mathbf{k}) = t \sum_i \exp(i\mathbf{k}\cdot \mathbf{a}_i)$ with $\mathbf{a}_i$, $i=1,2,3$ the connecting vectors from a given site to its three nearest neighbors and $\sigma_x,\sigma_y$ Pauli matrices.}, and $\mathbf{b}_i$, $i=1,2,3$ denote the three different types of vectors connecting a site with its next-nearest neighbors. At $t^{\prime} = \pm t_{\text{crit}} = \pm \Delta/3\sqrt{3}$, the gap closes at one of the Dirac points (i.e., they become massless and sources of Berry curvature) and the system undergoes a topological phase transition. Different from the 2D SSH model, the topological invariant is given by the Chern number, which can be calculated as $\mathcal{C}=-1,0,1$ for $t^{\prime}<-t_{\text{crit}}, -t_{\text{crit}}<t^{\prime}< t_{\text{crit}}, t^{\prime}>t_{\text{crit}}$, respectively.
Using ultracold fermionic atoms, the Haldane model has been successfully realized and explored experimentally~\cite{Jotzu2014}. 

We start by analyzing plasmonic responses in Haldane nanoribbons with armchair edges (aHNRs), i.e., finite strips with periodic boundary conditions in $y$-direction and armchair edges at the sides.
\begin{figure}
\centering
\includegraphics[width=1\linewidth]{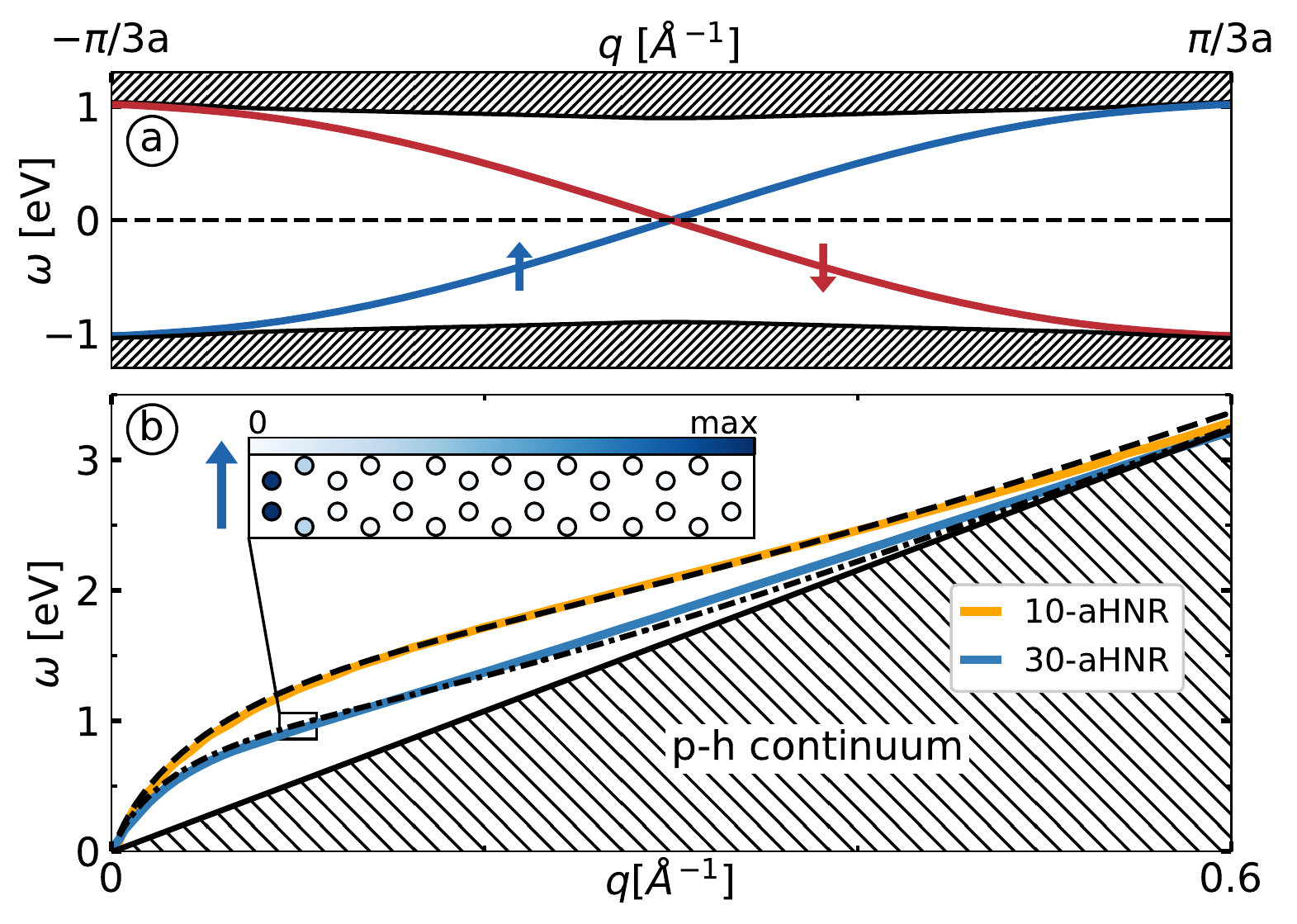}
\caption{(a) HNR band structure. Up (down) moving chiral edge states localized on the left (right) side of the ribbon are illustrated in blue (red) color. Densely shaded areas correspond to bulk bands. We set $t=2.7$ eV, $t' = 0.2t$, $\delta=0.2t$ and $\mu=0$ eV. (b) Up moving edge plasmon dispersion for HNRs with 10 (orange) and 30 (blue) atoms in the unit cell. The dashed area represents the p-h-continuum with upper edge $\omega_{ph} = v_F q$. The theoretical prediction Eq.~\eqref{eq:HNRdisp} with best fitting effective cross section ($\overbar{\sigma_C}\approx 3.9a, 6.2a$ for 10-aHNR and 30-aHNR, respectively) is shown in dotted/dashed-dotted black lines. Inset: representative induced charge density distribution in the unit cell at a chosen plasmonic excitation.}
\label{fig:ribbon}
\end{figure}
Fig.~\ref{fig:ribbon}~(a) shows the band structure of an aHNR in dependence of its 1D momentum $q=q_y$. The form of the unit cell is shown in the inset of Fig.~\ref{fig:ribbon}~(b), resulting in a unit cell length of $L=3a$. The chiral edge states emerging in non-trivial phases are localized only at one side of the HNR and propagate in opposite directions, following the handiness of the complex next-nearest neighbor interactions. We numerically find that the Fermi velocity of the chiral edge modes around $q=0$ is given by $v_F= 2t$, and is thus independent of the parameters $t'$ and $\Delta$. In order to make theoretical predictions of the plasmonic response, we approximate the bands as fully independent, with energy dispersions $E_{k,\pm} = \pm v_F k $, where the + (-) modes correspond to unidirectional up (down) propagation of the edge modes.
With the usual approximation of uniform intra-band correlations, the intra-band contributions to the polarization function can straightforwardly be evaluated,
\begin{flalign}
    \Pi^0_{\text{intra}, \pm} (\omega, q) &= \frac{g_s}{2\pi} \int_{\text{BZ}} dk \frac{\Theta(E_{k,\pm}<0)- \Theta(E_{k+q,\pm}<0)}{\omega + i\eta -v_F|q|} \\
    &= \frac{1}{\pi} \frac{|q|}{\omega + i\eta - 2t |q|}, \nonumber
\end{flalign}
which in RPA results in the plasmon dispersion,
\begin{equation}
    \omega(q) = \left\{ \frac{2e^2 K_0(\sigma_c |q|)}{\pi} + 2t \right\} |q|.
    \label{eq:HNRdisp}
\end{equation}
The cross section $\sigma_C$ here defines an inverse momentum scale which dictates the crossover region from logarithmic to linear dispersive behavior. 
There is, however, a subtlety that has to be be addressed, as the two chiral edge modes change localization and direction at the BZ boundaries and are therefore correlated with the bulk bands in this regime. This comes along with a strengthening of inter-band screening effects that cannot be neglected. For our theoretical predictions, we here make a rough phenomenological approach and account for the screening by using an effective $\overbar{\sigma_C}>\sigma_C$ in Eq.~\eqref{eq:HNRdisp}, which enlarges the overall Coulomb scattering cross section and hence qualitatively captures the delocalization effects at the zone boundary.
In our numerical calculations\footnote{In the case of ribbon structures, where periodic boundary conditions are assumed only in one direction, the (matrix like) polarization function $[\boldsymbol{\Pi}^0(\omega,q)]_{\sigma \sigma'}$ is given by \begin{equation}
\begin{split}
[\boldsymbol{\Pi}^0(\omega,q)]_{\sigma \sigma'} = \frac{g_s}{V} \sum_{k, l, l'} \frac{n_F(E_{l}^{k})- n_F(E_{l'}^{k+q})}{\omega+  i\eta + E_{l}^{k} -E_{l'}^{k+q}} \\  \psi_{l \sigma}^{k} \psi_{l \sigma'}^{k*} \psi_{l' \sigma}^{k+q*} \psi_{l' \sigma'}^{k+q}, \end{split} \end{equation} where V is the length of the unit cell in direction of periodic boundary conditions, $E_l^k$ is the band energy of band $l$ at momentum $k$, and $\psi_{l \sigma}^k$ represents entry $\sigma$ (i.e., the sub-lattice index) of the wave-function at momentum $k$ of band $l$.}, we fix $\sigma_C=a$. However, note that the concrete choice of the Coulomb cross section is only of minor importance for the here considered ribbon sizes if all inter-band contributions are accounted for, as is the case in the numerical evaluation of Eq.~\eqref{eq:Pi0}. Fig.~\ref{fig:ribbon}~(b) shows the computational results of the plasmon dispersion for two different ribbon widths with 10 and 30 sites per unit cell. By comparing them to the theoretical approximation Eq.~\eqref{eq:HNRdisp} using the {\it best fitting} effective cross section $\overbar{\sigma_C}$, we see how the numerical results including all inter-band screening and overlap effects follow the rough phenomenological prediction over a wide range of momenta surprisingly well. In~\ref{sec:ApxB}, the inter-band screening effects are analyzed and elaborated in more detail. Note that we expect measurable edge plasmons only for relatively small ribbon widths, as larger HNR sizes lead to {\it overscreening} of the edge modes and hence quickly lock them to the upper edge of the p-h continuum. For this reason, it is desirable to reduce screening effects from the bulk bands, which could e.g. be realized by tuning the system deeper into the topologically non-trivial phase via a strengthening of the complex next-nearest neighbor hopping inducing magnetic fields. Due to the chirality of the mid-gap topological edge states, the plasmons inherit uni-directory. For positive momenta (hence, up-moving plasmons), the induced charge density distributions are strongly localized on the left side for our choice of $t'$, see the inset of Fig.~\ref{fig:ribbon}~(b). For $q<0$, the quasi-1D plasmons move downward and are localized on the right side of the HNR. The topological phases $\mathcal{C}=\pm 1$ are related to each other only by a flip of the propagation direction of the single-particle and collective modes. 

Due to the symmetry of the edges, both up- and down-moving plasmon branches have identical dispersions. Including symmetry breaking adatoms on one edge leads to avoided level crossings and in turn results in different Fermi velocities for the two edge modes. We therefore expect a certain control of the dispersions for $q \lessgtr 0$ via edge manipulation, as already explored in the single particle picture in Chern insulator nanoribbons \cite{Malki2017}. 

\begin{figure}
\centering
\includegraphics[width=0.95\linewidth]{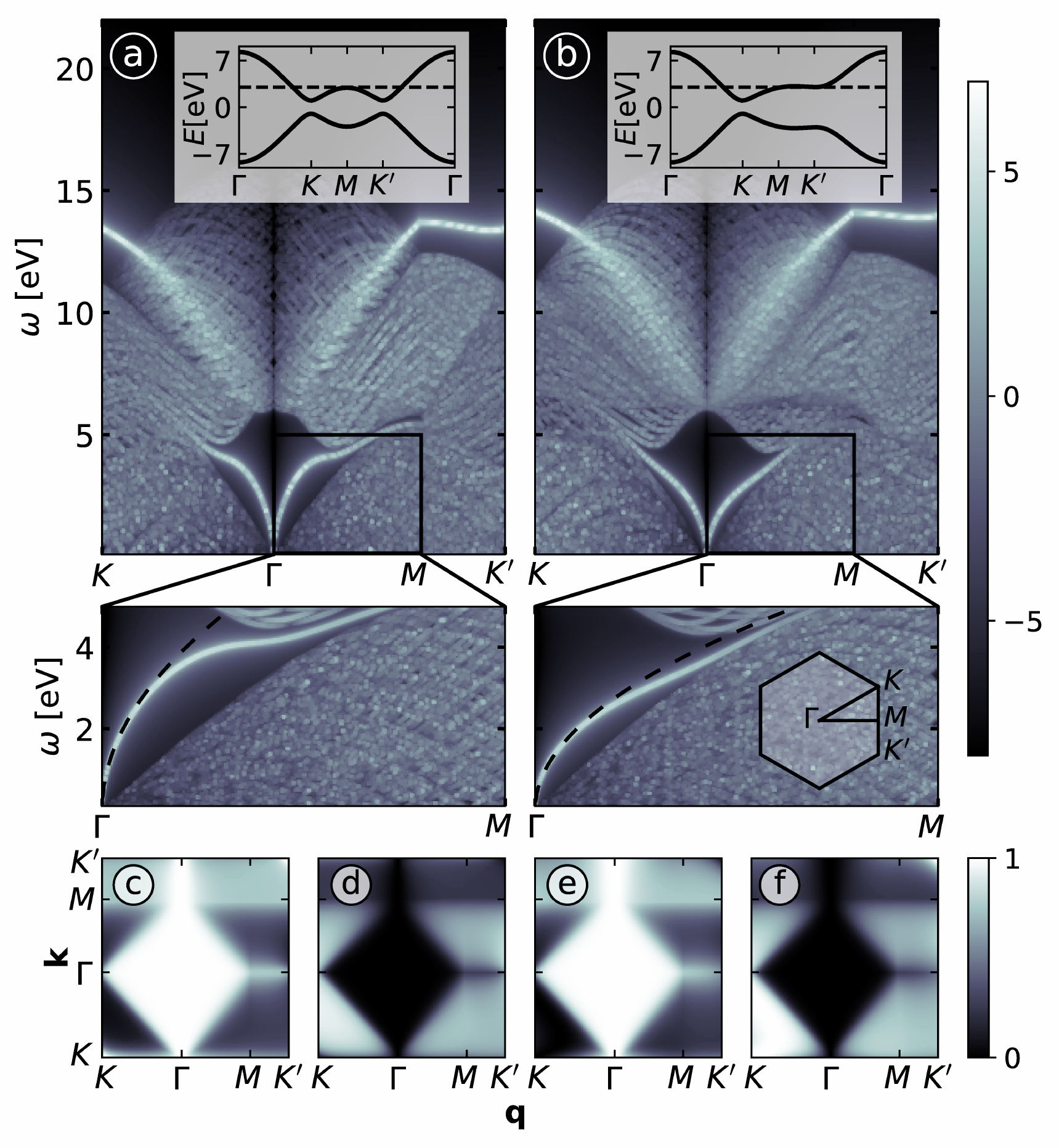}
\caption{Electron energy loss spectra of the Haldane model for (a) $\mathcal{C}=0$ ($t^{\prime}=0$) and (b) $\mathcal{C}=1$ ($t^{\prime}=2 t_{\text{crit}}$) with $\Delta=1, t=2.7$ eV and $d=1.42$ $\angstrom$. Insets: Single-particle energy dispersions (solid lines) and chemical potential $\mu=3$ eV (dashed line). Zooms below (a) and (b) include the analytical low $q$ expansions (black dashed lines) for intra-band plasmons. (c) and (d): inter- and intra-band overlap functions for $\mathcal{C}=0$, respectively, along high-symmetry paths for both $\mathbf{k}$ and $\mathbf{q}$. (e) and (f): the same for $\mathcal{C}=1$.}
\label{fig:haldane_plasmons}
\end{figure}
Let us finally discuss bulk plasmons in the doped Haldane model, focusing on differences in different topological phases. Numerical results of the electronic energy loss spectra are shown in Figs.~\ref{fig:haldane_plasmons} (a) and (b) for $\mathcal{C}=0$ and $\mathcal{C}=1$, respectively. Examining the low energy collective modes, we find that for $\mathcal{C}=1$, the plasmon dispersion is softened when comparing it to the trivial phase while keeping the chemical potential constant. For the energetically higher plasmonic branch, although the total energy of the non-trivial phase plasmon is slightly enhanced, the bandwidth remains indistinguishable from the $\mathcal{C}=0$ high-energy plasmon. Changing the chemical potential into the gap and/or increasing the next-nearest neighbor hopping strength to move the system further into the $\mathcal{C}=1$ regime does not fundamentally change these observations. Analyzing the intra- and inter-band overlap functions, Figs.~\ref{fig:haldane_plasmons} (c)-(f), we see that they barely vary when switching from $\mathcal{C}=0$ to $\mathcal{C}=1$. The term dominating a variation of the plasmon dispersion is hence identified as the change of single particle band structure details, illustrated in the insets of Fig.~\ref{fig:haldane_plasmons}. Indeed, we find that the low-$q$ expansion of the gapless plasmon mode is given by (\ref{sec:ApxC}),
\begin{equation}
    \omega^2(q) = \frac{2e^2 \big[ \mu - \big(\Delta^2+(3\sqrt{3}t^{\prime})^2\big)/\mu \big]}{\kappa} q, 
\end{equation}
which is included in the lower parts of Figs.~\ref{fig:haldane_plasmons} (a) and (b). The squared energy dispersion in the gapped system is hence quadratically softened by the gap parameter and the next-nearest neighbor hopping, in contrast to a linear decrease when simply lowering the doping level in conventional graphene, whose low energy plasmon mode is given by $\omega^2(q) =2e^2 \mu q/\kappa$~\cite{Wunsch2006, Hwang2007}~(see also \ref{sec:ApxC}).
%
%
%
%
\section{Conclusions}
\label{sec:Conc}
Using complementary real and momentum space approaches, we have examined the plasmonic excitations arising from long-range Coulomb interactions in two-dimensional models of topological insulators, and found several collective phenomena of interest.
First, we found that gapless, highly localized plasmons emerge in non-trivial phases, whose dispersion we analyzed for two topological insulator models. In the 2D SSH model, we observed collective modes that are localized on the boundaries and behave like bulk plasmons in a 1D system, with a Coulomb cross section of about the size of the lattice spacing. This suggests quasi-1D collective physics in a 2D material, coming with an additional layer of protection due to the underlying non-trivial topology. Furthermore, the (non-chiral) edge plasmons were shown to be stable against bulk screening effects, suggesting their existence independent of the system size. In the Haldane model, chiral, uni-directional localized plasmons were found, which were, however,  observable only in thin ribbons, due to the chiral modes' susceptibility to inter-band oversscreening.
The discussed tunability via edge manipulation opens future research questions regarding the control of quasi-1D plasmons in HNR architectures.
Moreover, it is of high interest to investigate the stability of the edge plasmons against disorder.
On the other hand, our analysis of bulk 2D SSH excitation spectra predicts strong dispersion hardening and softening when tuning the system between dimerized and anti-dimerized regimes, controlled by the intra- and inter-band wave function correlations. The strong enhancement of inter-band overlaps and plasmonic energies can be used to access regimes where Landau damping is entirely quenched, thus enabling applications based on dissipationless light-matter coupling. Bulk plasmons in the Haldane model, in contrast, were shown to almost entirely be controlled by single-particle band structure details when changing the system's topological phase, which we investigated analytically for small momentum transfers. 

\textit{Acknowledgements.---} We would like to thank Stefan Wessel, Hubert Saleur, Manfred Sigrist, Masao Ogata, and Ammon Fischer for useful discussions. This work was supported by the US Department of Energy under grant
number DE-FG03-01ER45908. The numerical computations were carried out on the University of Southern California and RWTH Aachen University High Performance Supercomputer Clusters.

\Apix
\section{}
\label{sec:ApxA}
\begin{figure}
\centering
\begin{subfigure}
  \centering
  \includegraphics[width=8.cm]{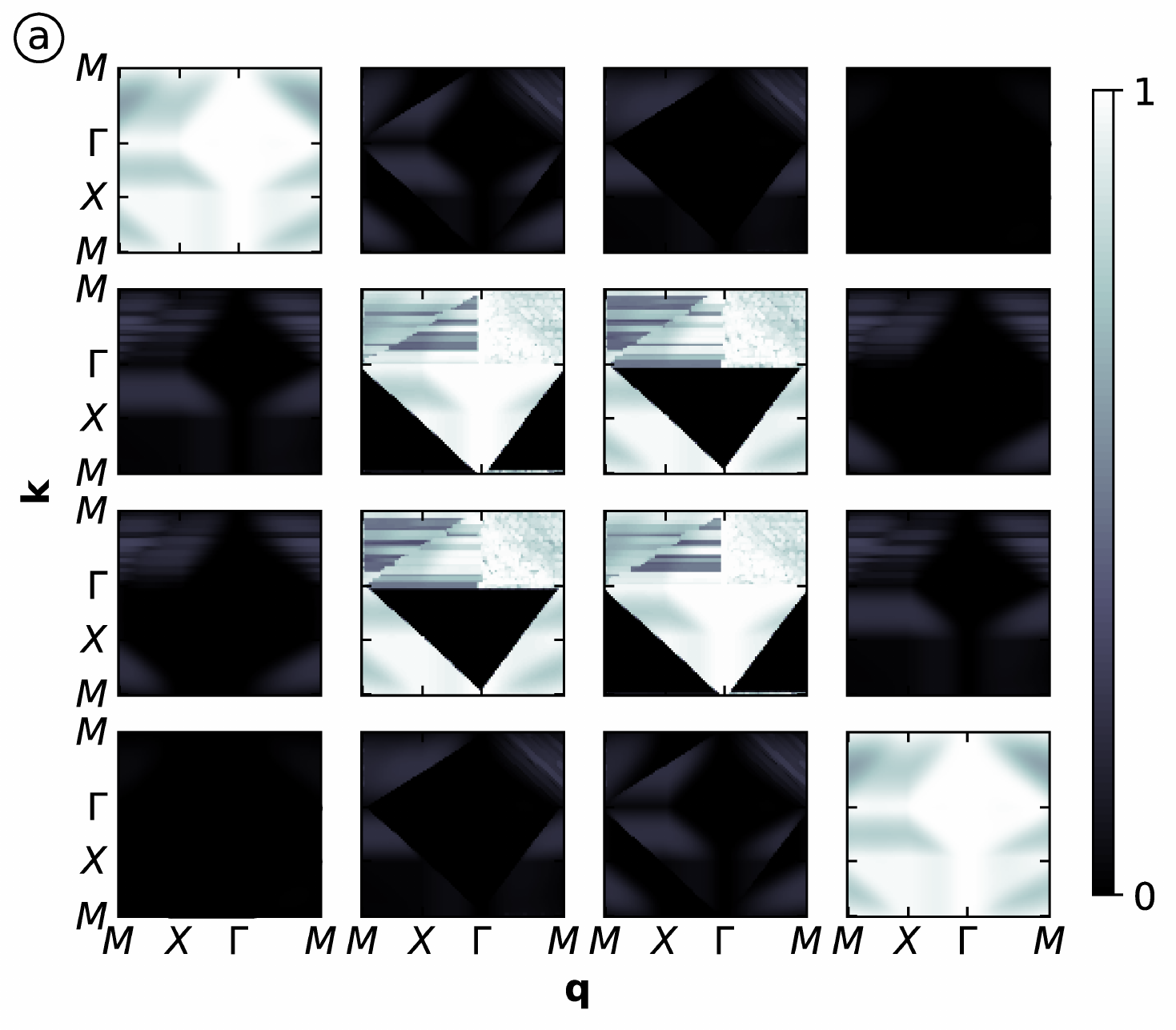}
\end{subfigure}%
\begin{subfigure}
  \centering
  \includegraphics[width=8.cm]{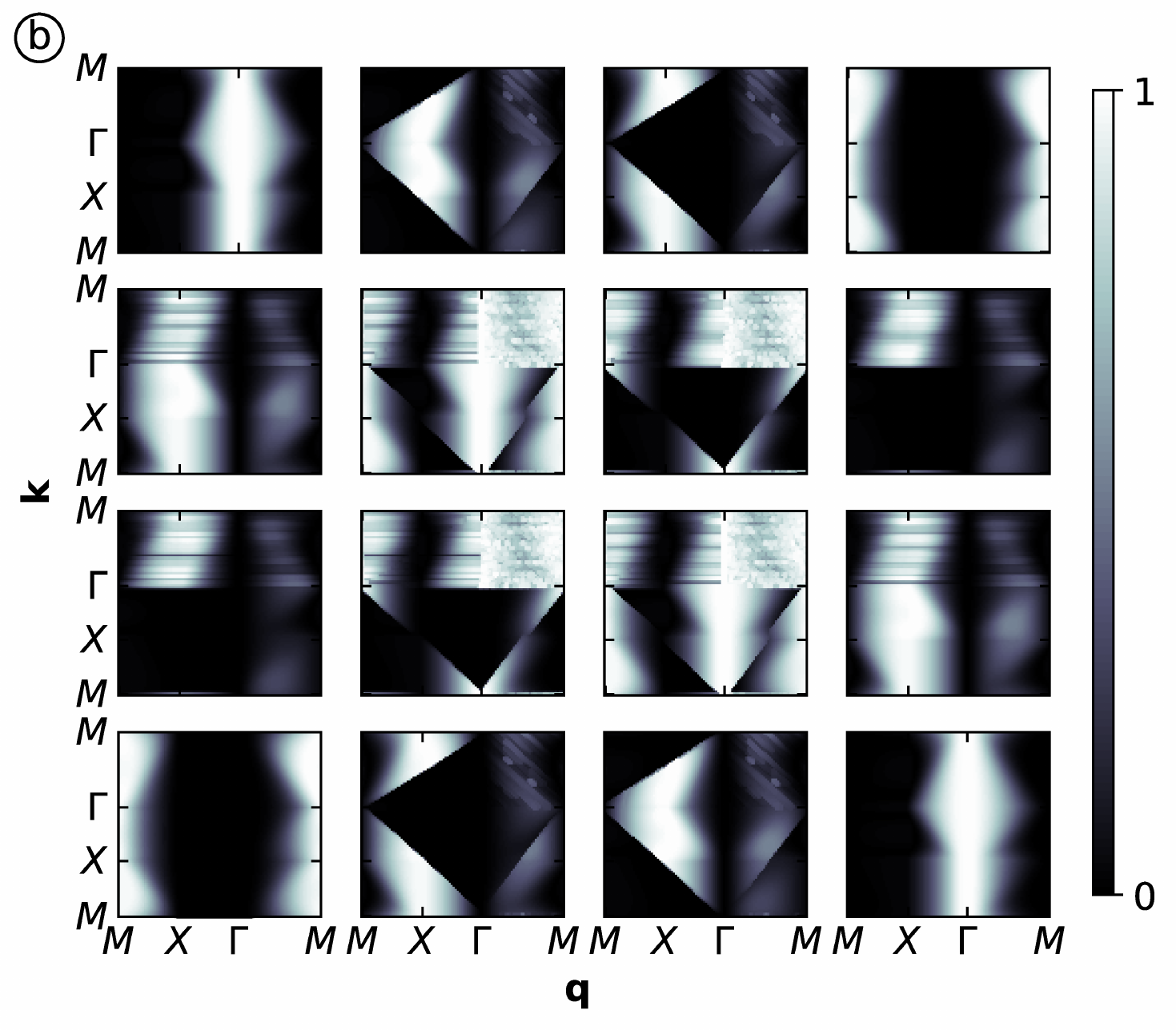}
\end{subfigure}
    \caption{All 16 overlap functions of the four-band two-dimensional SSH-model for (a) the dimerized and (b) the anti-dimerized regime. Intra-band overlaps ($l=l'$) are displayed on the diagonal, whereas inter-band form factors ($l \neq l'$) fill the off-diagonal elements of the grids. A path along the high symmetry points $M,X,\Gamma,M$ is chosen for both $\mathbf{k}$ and $\mathbf{q}$ to illustrate the overlap functions. Between $\Gamma$ and $M$ we find strong numerical noise if $l=p_x,p_y \vee l'=p_x,p_y$ due to the degeneracy of the $p$ bands in this region.}
    \label{fig:2dssh_overlap}
\end{figure}

The overlap functions, often also called coherence or form factors,
\begin{equation} F_{\mathbf{k}, \mathbf{k+q}}^{l  l'} = |\braket{\Psi_{\mathbf{k}, l}|\Psi_{\mathbf{k+q},l'}}|^2,
\end{equation} are identified in the main text as the only actors influencing the plasmonic dispersion in the 2D SSH model when switching between the relevant regimes D$\leftrightarrow$AD. All 16 overlap functions for chosen values of momentum $\mathbf{k}$ and transferred momentum $\mathbf{q}$ along high-symmetry points of the BZ (for $\delta=0.1L$ as in the main text) are shown in Fig.~\ref{fig:2dssh_overlap} (a) and (b) for the dimerized and anti-dimerized regime, respectively. Here, intra-band (inter-band) overlap functions are shown as diagonal (off-diagonal) elements. As a sanity check, note that for $\mathbf{q}=0$, i.e., at the $\Gamma$ point on the $\mathbf{q}$-axis, the intra-band (inter-band) overlap is one (zero) for all $\mathbf{k}$, as the hermiticity of the Hamiltonian demands. Generally, one can observe that the overlap of inter-band (intra-band) wave functions is greatly enhanced (suppressed) in the AD phase, Fig.~\ref{fig:2dssh_overlap} (b), compared to its analogue in the D regime, Fig.~\ref{fig:2dssh_overlap} (a). 
For the inter-band contributions, we hence observe that single particle transitions of high energy difference are enhanced (or, strictly speaking, less suppressed) in the AD phase and thus lead to significantly larger contributions to the polarization function when summing over $\mathbf{k}$ for a given $\mathbf{q}$. On the other hand, low energy intra-band transitions are suppressed in AD regimes, resulting in smaller contributions to $\Pi^0(\omega, \mathbf{q})$. This leads to the observed hardening (softening) of the high energy (low energy) plasmon mode, which is governed by inter-band (intra-band) transitions. In the dimerized phase, we further see that only inter-band transitions between $p$ bands play a significant role, such that one would expect the upper energy plasmon mode to be further softened when tuning the chemical potential to lie inside the gap between the $p_y$ and $d_{xy}$ band, hence prohibiting single-particle transitions between the two $p$ bands. The high-energy plasmon modes in AD phases, on the other hand, are not expected to alter by a considerable amount due to strong contributions coming from all other inter-band transitions. We confirmed this numerically.

\section{}
\label{sec:ApxB}

\begin{figure}
\centering
\includegraphics[width=1\linewidth]{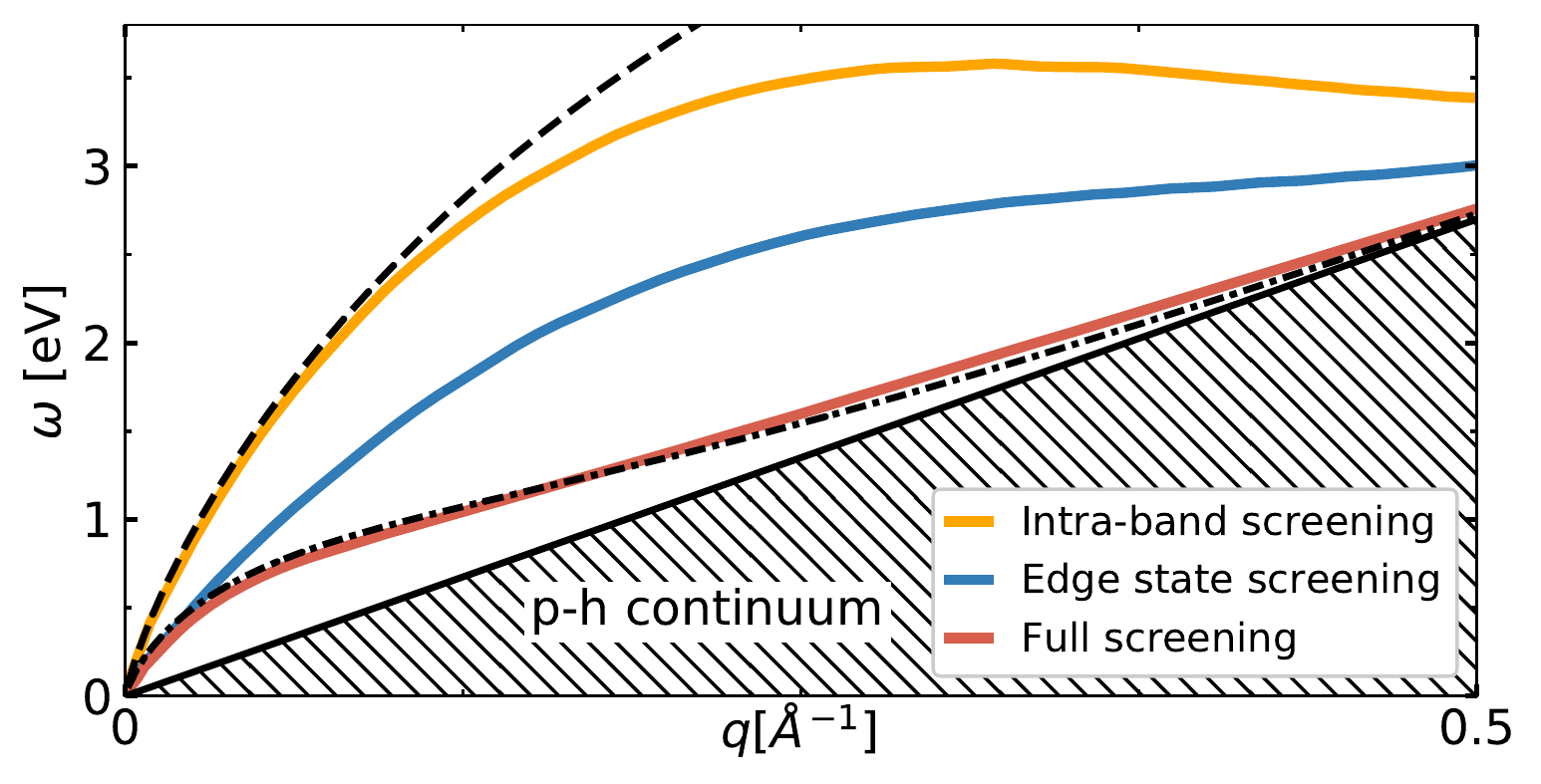}
\caption{Screening properties of a 30-aHNR. Plasmonic responses using Eq.~\eqref{eq:Pi0} are shown while taking into account (i) only intra-band contributions with energy $E_{k,+}$ (yellow line). (ii) screening from both chiral edge state bands (blue), and (iii) taking into account the full inter-band screening (red). The low-$q$ expansion Eq.~\eqref{eq:HNRdisp} for case (i) is displayed with a black dotted line. Increasing the Coulomb cross section to $\overbar{\sigma_C}$ to effectively describe screening effects results in the black dashed-dotted line.}
\label{fig:ribbon_screening}
\end{figure}

Here, we  analyze the inter-band screening effects for a 30-aHNR. Fig.~\ref{fig:ribbon_screening} illustrates the chiral plasmon dispersion when taking account succeedingly more inter-band contributions to the polarization function. 
Only considering intra-band screening from the chiral edge band with band energy $E_{k,+}$ results in the yellow dispersion. The theoretical prediction Eq.~\eqref{eq:HNRdisp} based on solely intra-band contributions is added with a black dashed line (in this case, $\sigma_C = \overbar{\sigma_C} = a$ in Eq.~\eqref{eq:HNRdisp}). For small momentum transfers, there is good agreement with the numerics. For larger momenta, the overlap function term in the polarization function softens the dispersion, consistent with the delocalization of the chiral edge state band at the borders of the BZ. Increasing the inter-band screening contributions to $\Pi^0(\omega,q)$ further softens the quasi-1D plasmon dispersion, as seen in Fig.~\ref{fig:ribbon_screening} for chiral edge state screening (taking into account both mid-gap bands) in blue and full bulk band screening in red. Introducing an effective enlarged Coulomb scattering cross section $\overbar{\sigma_C}$ in Eq.~\eqref{eq:HNRdisp} can reproduce the fully screened dispersion over a large range of momenta with surprising accuracy. We would like to clarify that this correspondence even for larger momentum transfers should, however, not be viewed as a general statement for chiral plasmons in Chern insulators, but does seem to effectively capture the inter-band screening effects in this particular setting.
\section{} 
\label{sec:ApxC}
Here, we present an analytical approximation of the low-energy $\sqrt{q}$ bulk plasmonic mode in the Haldane model. Let us start with the standard tight binding model for graphene, for which
\begin{equation}
    F_{\mathbf{k}, \mathbf{k+q}}^{l l'} = \frac{1}{2} \{ 1+ l l' \cos(\phi_{\mathbf{k}} - \phi_{\mathbf{k+q}}) \},
\end{equation}
where $\phi_{\mathbf{k}} = \text{arg}\{h(\mathbf{k})\}$.
For finite electronic doping $\mu>0$, the low energy plasmon mode is formed by intra-band transitions in the conduction band, for which the overlap function can be approximated by $F_{\mathbf{k}, \mathbf{k+q}}^{l= l'=+} = 1 + \mathcal{O}(q^2)$. Denoting the conduction band energies by $E_{\mathbf{k}}$ and expanding the Lindhard term in the bare polarization function in $\omega \gg |E_{\mathbf{k}}- E_{\mathbf{k+q}}|$, one finds
\begin{equation}
\begin{split}
    \Pi^0_{\text{intra}} &\approx \frac{g_s}{\omega^2} \sum_{\mathbf{k}} n_F(E_{\mathbf{k}}) \{ E_{\mathbf{k+q}} - E_{\mathbf{k-q}} -2E_{\mathbf{k}} \} \\
    &\approx \frac{g_s}{\omega^2} \sum_{\mathbf{k}} n_F(E_{\mathbf{k}}) (\mathbf{q}\cdot \nabla_{\mathbf{k}})^2 E_{\mathbf{k}}.
\end{split}
\end{equation}
Approximating the conduction band as $E_{\mathbf{k}} = v_F k$ throughout the whole BZ and accounting for the additional valley degeneracy factor $g_v = 2$, one finds $(\mathbf{q}\cdot \nabla_{\mathbf{k}})^2 E_{\mathbf{k}} =v_F \sin^2(\alpha) q^2/k$, with $\alpha$ being the angle between $\mathbf{q}$ and $\mathbf{k}$. Hence,
\begin{equation}
    \Pi^0_{\text{intra}} \approx \frac{4v_F q^2}{(2\pi)^2\omega^2} \int_0^{2\pi} \sin^2(\alpha) d\alpha \int_{0}^{k_F} dk = \frac{\mu q^2}{\pi \omega^2},
\end{equation}
which within RPA results in the well known low-$q$ plasmon square-root dispersion for graphene \cite{Wunsch2006, Hwang2007},
\begin{equation}
    \omega^2(q) = \frac{2e^2 \mu}{\kappa} q.
\end{equation}
When introducing a staggered sublattice potential and complex next-nearest neighbor hoppings to the Hamiltonian, the valley degeneracy is broken and the (now gapped) system can be approximated by two Dirac cones with a corresponding mass term
\begin{equation}
    E_{k,\pm}^P = \pm \sqrt{\Delta_P^2 + (v_F k)^2}, 
\end{equation}
where $\Delta_{P} = \Delta \pm 3\sqrt{3}t^{\prime}$ at $P=K'$ (+) and $P=K$ (-). Again keeping the chemical potential inside the conduction band and focusing on the intra-band transitions ($\rightarrow F_{\mathbf{k}, \mathbf{k+q}}^{l=l'=+} = 1 + \mathcal{O}(q^2)$), we find that
\begin{equation}
    (\mathbf{q}\cdot \nabla_{\mathbf{k}})^2 E_{k,+}^{P} = \frac{(v_F q)^2}{E_{k,+}^P} - \frac{(v_F q)^2(v_F k)^2}{\big(E_{k,+}^P\big)^3} \cos^2(\alpha).
\end{equation}
The polarization function hence reads
\begin{gather}
\begin{split}
    \Pi^0_{\text{intra}} \approx \frac{2 q^2}{(2\pi)^2\omega^2} &\sum_{P=K,K'} \int_0^{2\pi} d\alpha \int_{0}^{v_F k_F^P} dk \\ &\left\{ \frac{k}{\sqrt{\Delta_P^2 + k^2}} - \frac{k^3}{\sqrt{\Delta_P^2 + k^2}^3}\cos^2(\alpha) \right\} \end{split} \\ = \frac{\mu - \big[\Delta^2+(3\sqrt{3}t^{\prime})^2\big]/\mu }{\pi \omega^2} q^2 \nonumber, \end{gather} 
where $k_F^P$ is defined such that $\mu = \sqrt{\Delta_P^2 + \big(v_F k_F^P\big)^2}$. Within RPA, this results in the plasmonic energy dispersion,
\begin{equation}
    \omega^2(q) = \frac{2e^2 \big[ \mu - \big(\Delta^2+(3\sqrt{3}t^{\prime})^2\big)/\mu \big]}{\kappa} q.
\end{equation}
\bibliographystyle{apsrev4-1}
\bibliography{plasmons_2D_topo}
\onecolumngrid
\end{document}